\documentclass[aps,prd,preprint,groupedaddress,amssymb,onecolumn]{revtex4-1}
\usepackage[utf8]{inputenc}
\usepackage[breaklinks]{hyperref}
\usepackage{graphicx}
\usepackage{natbib}
\usepackage{outlines}
\usepackage{hyperref}

\usepackage{indentfirst}
\usepackage{amsmath}
\usepackage{xfrac}
\usepackage{subfigure}

\usepackage[usenames,dvipsnames]{pstricks}
\usepackage{epstopdf}
\usepackage{pst-grad} % For gradients
\usepackage{pst-plot} % For axes

\newsavebox{\astrutbox}
\sbox{\astrutbox}{\rule[-5pt]{0pt}{20pt}}

\usepackage{ulem}
\newcommand\DEL[1]{\sout{}}     % suggested deletion in text
\def\ADD#1{{\textcolor{black}{#1}}}    % suggested new text
\usepackage{graphicx}
\usepackage{color}

\begin{document}

\title{Scale effects in internal wave attractors}


\author{C. Brouzet$^1$, I.N. Sibgatullin$^{1,2}$, E.V. Ermanyuk$^{1,3,4}$, S. Joubaud$^1$, T. Dauxois$^1$}
\email[]{Thierry.Dauxois@ens-lyon.fr}
%\thanks{}
%\altaffiliation{1}
\affiliation{
	1. Univ Lyon, ENS de Lyon, Univ Claude Bernard, CNRS, Laboratoire de Physique, F-69342 Lyon, France \\
	2. Moscow State University, 119991, Institute for System Programming, 109004,
Shirshov Institute of Oceanology, 117997, Moscow, Russia\\
3. Lavrentyev Institute of Hydrodynamics, Novosibirsk 630090,
Russia\\
4. Novosibirsk State University, Novosibirsk 630090, Russia
}

\date{\today}

\begin{abstract}
As a necessary preliminary step toward geophysically significant extrapolations,
we study the scale effects in internal wave attractors in the linear and nonlinear regimes.
We use two geometrically similar experimental set-ups, scaled to factor 3,
and numerical simulations (a spectral element 
method, based on \ADD{the} Nek5000 open solver) for a range of parameters \DEL{which}\ADD{that} is 
typically accessible in laboratory.  
In the linear regime, we recover the classic\ADD{al} viscous scaling for the beam width, which 
is not affected by variations of the amplitude of the input perturbation. In the nonlinear regime,
we show that the scaling of the width-to-length ratio of the attractor 
branches is intimately related with the energy cascade from large-scale energy 
input to dissipation. We present results for the wavelength, amplitude and width of the beam
as a function of time and as a function of the amplitude of the forcing.

\end{abstract}

%\keywords{}

\maketitle

\section{Introduction}

Physical effects scale differently with size. In experimental fluid mechanics, the role of molecular viscosity is strongly exaggerated at the laboratory scale as compared to large-scale flows. This effect is particularly significant in geo- and astrophysical fluid mechanics owing to the size gap of several orders of magnitude between the real objects and their laboratory counterparts~\cite{SommeriaDidelle2009}. 
To unveil the key features of geo- and astrophysical flows, the modern laboratory experiments \DEL{put an accent onto}\ADD{emphasize the} exploration of strongly forced regimes (see e.g. ~\citep{GLS2016,LeBarsCefronLeGal2015,BSMPSE2006,VPBOP2010}). Under strong forcing, the purely viscous mechanism of momentum transfer by thermal molecular motion is replaced by momentum transfer due to vortices and/or waves. Accordingly, scaling laws and similarity relations clearly defined for laminar regimes are replaced by more complicated ones~\cite{BC1998}, which are closer to large-scale reality but still not readily extrapolated to geo- and astroscales~\cite{GLS2016}. 

Internal waves in a uniformly stratified fluid obey a highly specific \ADD{Boussinesq} dispersion relation \DEL{which}\ADD{that} reads $\sin \theta= \pm\,\Omega_0$, where $\theta$ is the angle between the phase (respectively group) velocity vector and the vertical {(resp. horizontal)} direction,  $\Omega_0=\omega_0/N$ is the forcing frequency normalized by the buoyancy frequency $N=[(-g/\bar{\rho})({\rm d}\rho/{\rm d}z)]^{1/2}$, with $g$ the gravity acceleration, $\rho (z)$ the density distribution along the vertical coordinate~$z$ and~$\bar{\rho}$ a reference value~\citep{MowbrayRarity1967}. {Note that the $z$-axis points upwards, as shown in Fig.~\ref{fig:setup}}. This dispersion relation does not contain any length scale and admits the wave propagation in form of oblique wave beams. In some cases, the cross-beam structure of internal waves can be prescribed by specific motion of a rigid boundary~\cite{MMMGPD2010}. However, more typically, the {\it a priori} unknown scaling of the cross-beam structure with the key parameters of the problem represents an issue of central interest. Indeed, in an ideal (nonviscous) uniformly stratified fluid, there is no mechanism preventing singular behavior for certain geometrical settings, with infinitely high energy density at characteristic lines. The ``healing" of these singularities requires certain assumptions on the energy-dissipation mechanism, which can either be linear (purely viscous) or involve highly nonlinear processes of vorticity generation, wave breaking, wave-wave and wave-current interactions, streaming, etc. 

An important case is precisely a consequence of the dispersion relation. Internal waves \DEL{have}\ADD{obey} a very specific reflection law: the angle between the internal-wave beam and the vertical must be conserved when the beam is reflected at a rigid boundary. This provides a geometric reason for strong variation of the beam width (focusing or defocusing) upon reflection at a slope~\cite{DauxoisYoung1999}. It is noteworthy that at nearly critical slopes (i.e. when the wave beam slope is close to the topographic slope), numerical simulations reveal the transition to turbulence~\cite{GayenSarkar2010}. The transition occurs at moderate values of the Reynolds number \DEL{which}\ADD{that}, however, are higher than those typically accessible in the laboratory setups. Thus, the local energy losses due to wave reflection at a slope in the ocean and in the laboratory are likely to be governed by different scalings. 

The role of molecular viscosity on the structure of internal wave beams  has been studied in some detail for oscillations of isolated bodies of simple geometry in a uniformly stratified fluid of infinite extent~\cite{HurleyKeady1997}. Due to viscous effects, the singularities along characteristic lines tangent to the surface of the body~\cite{Hurley1997} are replaced by interior shear layers merging  into a single beam evolving with distance toward a self-similar solution~\cite{ThomasStevenson1972}. \DEL{S}\ADD{The s}olution \ADD{of Hurley and Keady}~\cite{HurleyKeady1997} has been shown to hold experimentally with high accuracy at the laboratory scale~\cite{Sutherlandetal1999}. However, its extrapolation to oceanographic scales can hardly be justified: in the case of a body comparable to a significant bottom topography, the decay rate of internal waves is too low to dissipate the energy of internal tides at reasonable distance even if a significant turbulent viscosity (an order of magnitude higher than the molecular one) is used for estimates~\cite{VEF2011}. This simple example implies that, at the ocean scale, the decay of internal-wave energy is governed by an energy cascade with yet unknown scaling properties.

In closed two-dimensional fluid domains filled with linearly stratified fluid, the focusing of internal waves usually prevails, leading to convergence of internal wave rays toward closed loops, the internal wave attractors~\cite{MaasLam1995}. Owing to similarity of the dispersion relation, a similar effect \ADD{can} occur\DEL{s} for inertial waves\DEL{ in axisymmetric geometrical settings}, where a special attention has been drawn to spherical liquid shells due to their relevance to the structure of celestial bodies~\cite{Stewartson1971,Stewartson1972}. The concept of interior shear layers has been successfully applied to regularize the singularities at characteristic lines, leading to \DEL{final}\ADD{finite} energy density at the attractor loops both in the cases of internal and inertial waves~\cite{RGV2001,RVG2002,HBDM2008,GSP2008}. Since attractors are usually considered in a closed fluid domain, there should be a balance between the energy injection into the domain at global scale and the dissipation at small scales. In this relation, the purely viscous dissipation in interior shear layers represents an important ``prototype problem", with possible extension to a variety of dissipative mechanisms~\cite{Ogilvie2005}. The solution of this problem yields a scaling for the equilibrium width of the attractor branches, which is set by the balance between the geometric focusing and a particular dissipative mechanism or a combination of such mechanisms. In the linear regime, the experimental results in nominally two-dimensional problems can be significantly ``contaminated" by the effect of dissipation at lateral walls of the test tank, which can represent a non-negligible fraction of the total dissipation~\cite{BrouzetJFM2016}. The interplay of the viscous losses in interior boundary layers, at lateral walls, and zones of the wave-beam reflections is considered in~\cite{BeckebanzeJFM2017}, along with the associated scalings. In the nonlinear regime, the energy transfer to small scales in unstable attractors operates via a cascade of triadic interactions~\cite{SED2013,BrouzetEPL2016,BrouzetJFM2016,BrouzetJFM2017}. A qualitatively similar nonlinear regime has been revealed in~\cite{JouveOgilvie2014} for inertial wave attractors. The calculations presented in~\cite{JouveOgilvie2014} demonstrate that the width of attractor branches increases for larger forcing.

In the present paper, we explore the scale effect in internal wave attractors in the linear and nonlinear regimes, both experimentally and numerically, for a range of parameters \DEL{which is }typically accessible in laboratory.  The scaling of the width-to-length ratio of the attractor branches is intimately related with the energy cascade from large-scale energy input to dissipation. We consider this study as a necessary preliminary step toward geophysically significant extrapolations. The wave attractors are considered in a classic\ADD{al} trapezoidal geometric setting~\cite{MBSL1997}. We use the experimental setup described in ~\cite{SED2013,BrouzetEPL2016,BrouzetJFM2016,BrouzetJFM2017} and its scaled (to a factor 3) version to reveal the nonlinear scale effects. 
In view of the effects due to viscous losses at lateral walls~\cite{BeckebanzeJFM2017}, we performed also two-dimensional simulations, with a version of the computer code used in~\cite{BrouzetEPL2016,BrouzetJFM2016}, which demonstrate the  universality of results for the beam's width under appropriate scaling for a range of parameters typically accessible in experiments. Quantitatively similar results are expected for sufficiently wide test tanks. In Section~\ref{Experimental and numerical setups}, we describe the experimental setups and the numerical method. In section~\ref{Scaling in stable attractor}, we present the experimental results for stable wave attractors,
emphasizing interior shear layers. In Section~\ref{Scaling in unstable attractors: a signature of triadic resonance instability},
we consider {experimental and numerical} unstable attractors emphasizing the role of the triadic resonance instability in setting the wavelength and the amplitude.
Conclusions are presented in Section~\ref{Conclusion}.

\section{Experimental and numerical setups}\label{Experimental and numerical setups}

\subsection{Experimental setup}

\begin{figure}[b!]
	\begin{centering}
		\includegraphics[width=0.5\textwidth]{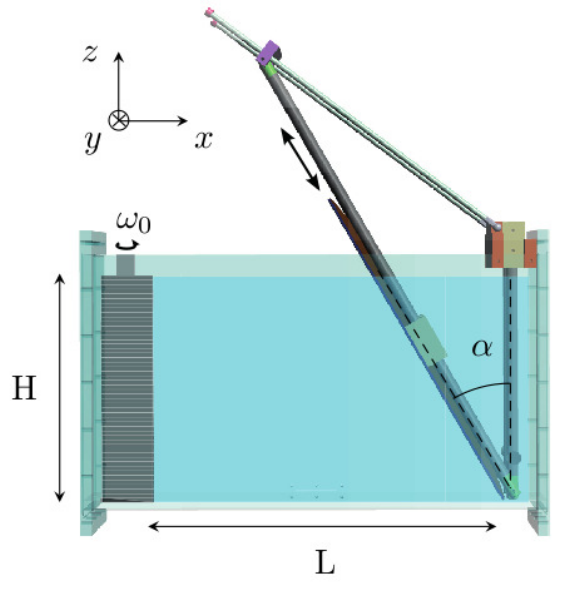}
		\caption{Geometric configuration of the experimental set-ups, with the wave generator prescribing the motion of the vertical wall, and the sloping wall inclined at an angle~$\alpha$ with respect to the vertical. The sloping wall delimits the trapezoidal domain with the working length $L$ and the depth~$H$.\label{fig:setup}}
	\end{centering}
\end{figure} 

In this work, we use two experimental setups. Both setups have the geometric configuration shown in Fig.~\ref{fig:setup}. The small setup~\cite{SED2013,BrouzetEPL2016,BrouzetJFM2016,BrouzetJFM2017} has the rectangular test tank of size length $\times$ height$\times$ width $= 80\times 42.5 \times 17.5~$~cm$^{3}$, with a typical working depth $H=30$~cm, while the large setup has the test tank of size $200\times 100 \times 17.4~$cm$^{3}$, with a typical working depth $H=90$~cm. The tanks are filled with an uniformly stratified fluid using the conventional double-bucket technique. Salt is used as a stratifying agent and the resulting density profile is linear.
 The sloping wall, which delimits the trapezoidal fluid domain of length~$L$ (measured along the bottom), is slowly inserted into the fluid after the end of the filling procedure.

In both setups, the input forcing is introduced  by a generator spanning the total fluid depth installed parallel to the vertical end wall. The time-dependent vertical profile of the generator is prescribed in the form of half-wave of a cosine function
\begin{equation}
\zeta(z,t)=a\sin(\omega_{0} t) \cos(\pi z/H),
\label{generator}
\end{equation}
where $a$ and $\omega_{0}$ are the amplitude and frequency of oscillations, respectively. In {the small} setup, we use the generator described in~\cite{GostiauxEF2007,MMMGPD2010,JMOD2012}, which reproduces the profile in discrete form by the horizontal motion of a stack of $47$ plates. The width of the plates is $3$~cm smaller than the width of the test tank due to the presence of supporting frame housing the plates. In {the large} setup, we use a generator, which reproduces the prescribed profile by motion of a flexible plate simply supported in the middle and clamped at the upper and lower ends to horizontal translating supports undergoing harmonic oscillations in anti-phase. The amplitude of these oscillations, $a$, can be easily varied with small increments in a sufficiently wide range (typically between $0$ and $6$~mm), to explore the nonlinear effects. The flexible plate is sealed at the sides to avoid water exchanges between the two faces of the plate. With this arrangement, the forcing is applied to the whole width of the fluid domain. 

The whole field measurement of the density-gradient perturbations is performed with the synthetic schlieren (SyS) technique~\cite{Sutherlandetal1999,DHS2000} from the field of local displacements of a background random dot pattern observed through the stratified fluid. The distortion of the dot pattern is recorded at the frame rate of $2$~fps by a computer-controlled video AVT (Allied Vision Technologies) Stingray camera with CCD matrix of $1388 \times 1038$ pixels or Pike camera with CCD matrix of $2452 \times 2054$ pixels, placed at a distance from the test tank {of} $165$~cm {for the small setup} and $250$~cm {for the large setup}. Conversion of images into the fields of density-gradient perturbations is performed using a cross-correlation PIV algorithm with subpixel resolution~\cite{FinchamDelerce2000} in combination with the basic relations for SyS~\cite{Sutherlandetal1999,DHS2000}. The temporal resolution of the measurements is (typically) around 20 fields per wave period. The spatial resolution of the field of the density-gradient perturbations is about $3$ and $9$~mm in each direction for the small and large setups, respectively. {\DEL{It}\ADD{This} is found sufficient to resolve the fine details of the internal wave field.} The subsequent processing allowing the separation of different components of the internal wave field is performed with the help of Fourier and Hilbert filtering~\cite{MGD2008} similar to~\cite{SED2013,BrouzetEPL2016,BrouzetJFM2016,BrouzetJFM2017}.   

In the present paper, we discuss mostly the experiments performed in the large tank in the linear and nonlinear regimes. Table~\ref{tab:series_exp} contains the forcing amplitude of the experiments, the regime observed  and the symbols used to plot the data. The experimental data obtained in the small tank, similar to those described in~\cite{SED2013,BrouzetEPL2016,BrouzetJFM2016,BrouzetJFM2017}, are used in Section III for cross-comparison with the data obtained in the large tank in the linear regime.  

\begin{table}[htb]
	\begin{center}
		\begin{tabular}{|c|c|c|c|}
			\hline
			Experiments & $a$~[mm] & Stability & Symbols \\
			\hline
			\hline
			$1$ & $0.7$ & Stable & Green pentagons\\
			\hline
			$2$ & $1.5$ & Stable & Magenta hexagons\\
			\hline
			$3$ & $2.2$ & Stable & Blue circles\\
			\hline
			\hline
			$4$ & $2.9$ & Unstable & Cyan triangles\\
			\hline
			$5$ & $3.7$ & Unstable & not plotted\\
			\hline
			$6$ & $4.4$ & Unstable & Red squares\\
			\hline
			$7$ & $5.1$ & Unstable & not plotted \\
			\hline
			$8$ & $5.8$ & Unstable & Black diamonds\\
			\hline
		\end{tabular}
		\caption{Experiments performed in the large tank, with always the same geometrical parameters: $H=92.3$~cm, $L=145.5$~cm, $\alpha=27.4^{\circ}$ and $\Omega_0=0.57$ leading consequently to $(d,\tau)=(0.34,1.81)$. {These notations are introduced in section~\ref{Scaling in stable attractor: theory}.} The symbols used to plot the wavelength, width and amplitude of the beam in Figs.~\ref{fig:etablissement_attracteur_stable},~\ref{fig:width_evolution} and~\ref{fig:wave_length_TRI} are detailed in the right column. Note that two experiments (numbers $5$ and $7$) have not been plotted for the sake of clarity.}
		\label{tab:series_exp}
	\end{center}
\end{table}

\subsection{Numerical setup}

For the numerical computations, we employ the spectral element 
method~\cite{FischerRonquist1994,Fischer1997,FischerMullen2001}. The 
numerical implementation of the method, based on \ADD{the} Nek5000 open solver, 
is known as a robust tool of direct numerical simulations of geo- and 
astrophysical flows involving highly nonlinear and non-trivial 
long-term dynamics (see e.g.~\cite{Favieretal2014,Favieretal2015}). In 
application to internal wave attractors, the method has been 
thoroughly tested by cross-comparison with the experimental 
results~\cite{BrouzetJFM2016}. In particular, the cross-comparison
demonstrated a very good qualitative and quantitative agreement 
between numerical experimental data, including the nonlinear regimes 
with well-developed instability.  The full system of equations being 
solved consists of the Navier-Stokes equation in the Boussinesq 
approximation, the continuity equation and the equation for the 
transport of salt. We impose the no-slip boundary condition at rigid 
walls and stress-free condition at free surface. Forcing is applied at 
the vertical wall by prescribing the profile of the horizontal 
velocity which reproduces the motion of the generator~\eqref{generator}. 

The meshes used in calculations for 3D modeling of
turbulent flows consist of 50 thousands to half-million elements. To
study 2D flows with moderate supercriticality, as in this paper, we
used meshes up to 5 thousand elements, with 8 to 10-order
polynomial decomposition within each element. Time discretization is
set to $10^{-4}$ to $10^{-5}$ of the external forcing period. We refer
the interested reader to~\cite{BrouzetJFM2016} for additional details.

The present paper reports the results of 2D numerical simulations for
a set of geometrically similar setups, which differ only in scale. The
2D \DEL{statement}\ADD{setting} allowed to exclude the energy losses due to friction at
the lateral walls~\cite{BeckebanzeJFM2017} and to study the effect of
triadic instability on scaling of the wave-beam width in isolation.

\section{Scaling in stable attractor: interior shear layers}\label{Scaling in stable attractor}

\subsection{Theoretical preliminaries}\label{Scaling in stable attractor: theory}

\begin{figure}[b!]
	\begin{centering}
		   \includegraphics[width=0.5\textwidth]{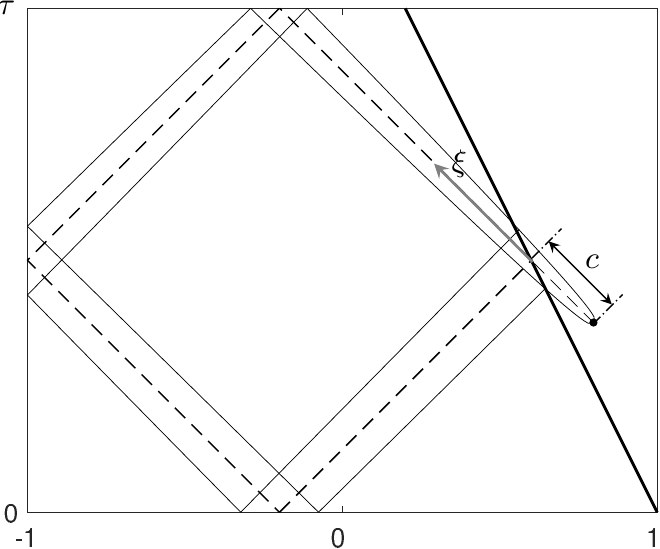}
		\caption{Ray tracing prediction (dashed line) for an attractor with $(d,\tau)=(0.2,1.6)$. The slope is represented by the thick solid black line. The virtual point source is indicated by the black dot, at the right of the slope. It emits a beam, whose width scales to the power $1/3$ with the distance from the source. Branches \DEL{are larger and larger}\ADD{grow wider and wider} until the focusing reflection on the slope. \ADD{The distance $c$ and the coordinate $\xi$ are indicated in the figure.}\label{fig:virtual_source}}
	\end{centering}
\end{figure}

The configurations of internal wave attractors in a fluid domain of particular geometry can be found by ray tracing~\cite{MaasLam1995,MBSL1997}. Plotting the rate of convergence of {the }wave rays toward the limiting cycles in \DEL{plane}\ADD{the} $(d,\tau)$ \ADD{plane}, where $d=1-(2H/L) \tan \alpha$ and  $\tau=(2H/L) \sqrt{1/\Omega_0^2-1}$ for the particular case of trapezoidal geometry~\cite{MBSL1997}, reveals \DEL{the }triangular domains corresponding to high convergence, which is typical for simple $(n,m)$-type wave attractors. Here, $n$ and~$m$ correspond to the number of ray reflections the attractor makes at the vertical and horizontal walls of the trapezoidal domain. In what follows, we restrict our attention to the simplest case of a $(1,1)$ attractor depicted in Fig.~\ref{fig:virtual_source}. The closed parallelogram (shown by dashed lines delimited by the trapezoidal boundary of the fluid domain in Fig.~\ref{fig:virtual_source}) represents a singular ``skeleton" given by ray tracing.  

Different models have been developed to explain the realistic structure of the attractor beams in a real viscous fluid~\cite{GSP2008,HBDM2008,JouveOgilvie2014}.  The beam formed by the branches of the attractor can be seen as a beam emitted by a virtual point source, located behind the focusing reflection~\cite{GSP2008,JouveOgilvie2014} as illustrated in Fig.~\ref{fig:virtual_source}. The beam then represents an interior shear layer, which has \DEL{the}\ADD{a} width $\sigma$ that scales as the distance from the virtual source to the power $1/3$, in agreement with self-similar solution~\cite{ThomasStevenson1972}. The distance $c$ between the focusing reflection and the virtual point source is set by this scaling law and the geometry of the attractor. Introducing $\xi$ as a coordinate along the ``skeleton" of the attractor measured from the focusing reflection, we have for the beam width 
\begin{equation}
\sigma(\xi) \propto (\xi+c)^{1/3}.
\label{eq:sigma}
\end{equation}
On the other hand, the beam widths before and after reflection are related to the focusing parameter $\gamma$ so that
\begin{equation}
\gamma=\frac{\sigma(L_p)}{\sigma(0)},
\label{eq:gamma}
\end{equation}
where $L_p$ is the perimeter of the attractor. Using equations~(\ref{eq:sigma}) and~(\ref{eq:gamma}), one has
\begin{equation}
c=\frac{L_p}{\gamma^3-1}.
\end{equation}
Following~\cite{GSP2008}, in our notations, we can write for the beam width normalized by the perimeter the following scaling 
\begin{equation}
\frac{\sigma(\xi)}{L_p} = C \left(1-\Omega_0^2\right)^{-1/6} \left(\frac{\nu}{N L_p^2}\right)^{1/3}\left(\frac{\xi+c}{L_p}\right)^{1/3}.
\label{eq:scaling_width}
\end{equation}
The implications of this scaling to geometrically similar setups and to geophysically signifiant scales are discussed below.

\subsection{Experiments with stable attractors in the linear regime}

\subsubsection{Definition of the wavelength, the width and the amplitude of the beam\label{branch_wavelength_width}}

\begin{figure}[b!]
	\begin{center}
		\includegraphics[width=0.85\linewidth,clip=]{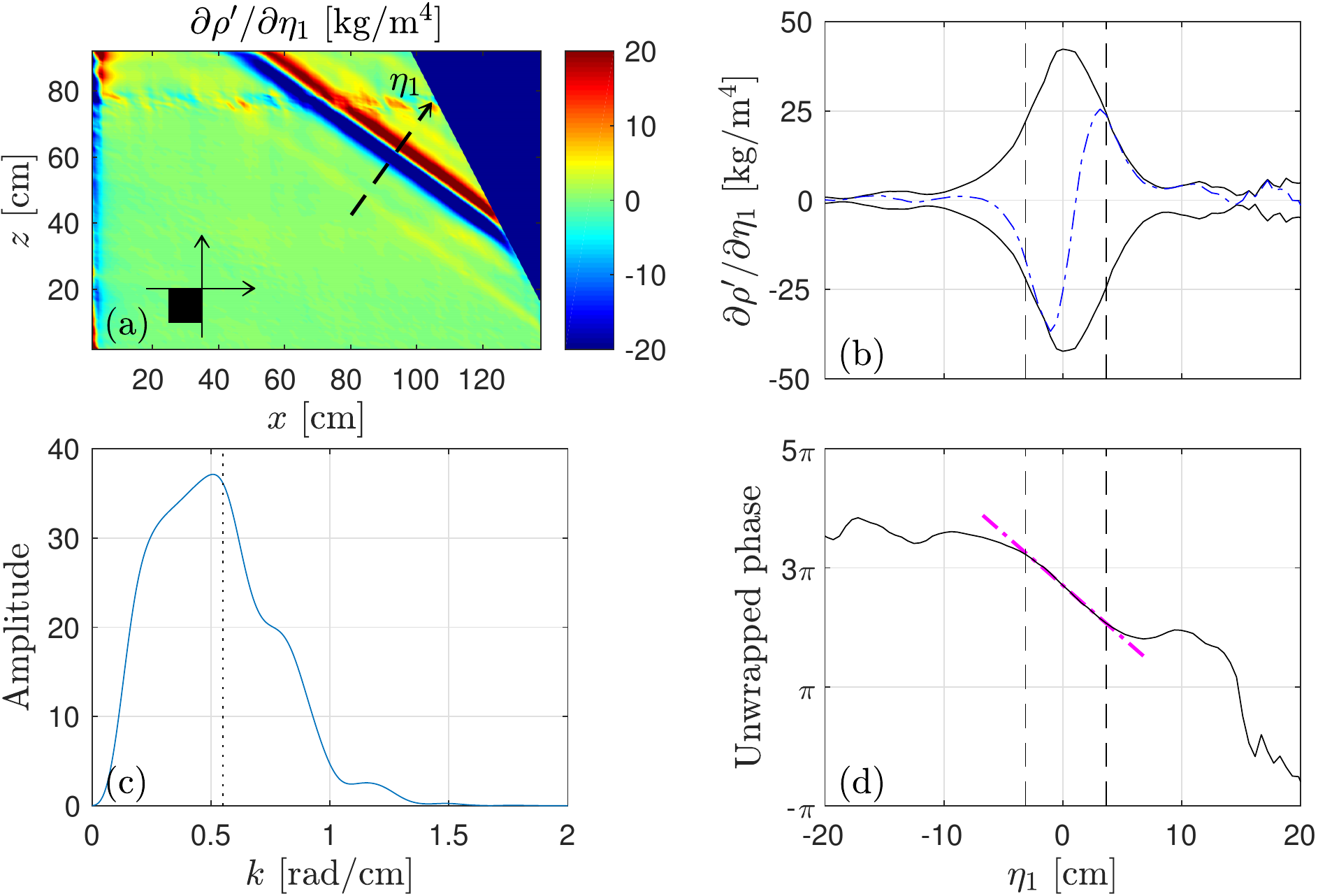}
		\caption{(a): \ADD{Real part of }$\partial \rho' / \partial \eta_1$ of a stable attractor observed with SyS during the steady state, after Hilbert filtering in frequency and space. The black square indicates where the \DEL{space }filtering in wave vector\DEL{s} \ADD{space} is performed. The cut made through branch $1$ is plotted as a dashed black {arrow}. (b):~$\partial \rho' / \partial \eta_1$ along the cut through branch~$1$. The real part is represented by a dash-dotted blue line and the modulus by the two black lines. (c): Fourier spectrum of the \DEL{real part of}\ADD{full complex signal} $\partial \rho' / \partial \eta_1$\DEL{ shown on panel~(b)}. The vertical dotted line shows the peak wave number found using the phase on panel~(d). (d): Unwrapped phase (black solid line) and fit of the central part (magenta dash-dotted line). The fit is performed between the two vertical dashed lines, on both panels (b) and (d). The distance between these two lines defines the width $\sigma$ of branch~$1$.\label{fig:explication_coupe}}
	\end{center}
\end{figure}

The wave number and the width of the attractor are measured using experiments performed in the small and large tanks with SyS as a visualization technique. The data processing is similar to~\cite{SED2013,BrouzetJFM2016,BrouzetJFM2017}. The different branches of the attractor are separated using Hilbert filtering in frequency and in space~\cite{MGD2008}. The branches are numbered from $1$ to $4$ in the direction of the energy propagation, with the branch $1$ starting at the focusing reflection. The results obtained for all branches are similar. Below we discuss in detail the measurements performed in branch $1$. The filtered horizontal and vertical gradient density fields are combined to the transverse density gradient field, $\partial \rho' / \partial \eta_1$, where $\eta_1$ is the transverse coordinate perpendicular to $\xi_{1}$ and  $\rho'$ the density perturbation with respect to the initial linear density profile. Th\DEL{is }\ADD{e real part of this} field is plotted in Fig.~\ref{fig:explication_coupe}(a). A cut, plotted as a dashed black line, is made through branch~$1$. Along this cut, one gets the real \ADD{part of the} signal, its modulus (amplitude) and also the phase. These quantities are plotted as a function of the distance~$\eta_1$, in Figs.~\ref{fig:explication_coupe}(b) and (d). The amplitude of the beam $|\partial \rho' / \partial \eta_1|$ is defined as the maximum of the modulus, at $\eta_1=0$. The phase is related to the wave-vector by $\vec{k}=-\overrightarrow\nabla{\phi}{=k\overrightarrow{e_\eta}}$, 
since the wave-vector is perpendicular to the beam. When unwrapping the phase, one can apply a linear fit in the vicinity of the maximum wave amplitude to estimate the slope\ADD{~\cite{BrouzetJFM2016,SED2013}}. This estimate yields the \ADD{wavelength $\lambda$, related to the }peak wave number $k_{\textrm{peak}}$\DEL{, }\ADD{. This quantity is measured in~\cite{HBDM2008} using the maximum of} \DEL{which is consistent with }the Fourier spectrum of the \DEL{real part of the}\ADD{full complex} signal, shown on panel (c). \ADD{As the maximum of the spectrum is close to the dotted line representing the wavenumber measured with the phase, both methods of measurements are consistent.} The width of the branches is defined as the width at half maximum \ADD{of the modulus of the signal.} \DEL{Consequently, the width is}\ADD{It corresponds to} the distance between the two vertical dashed black lines, surrounding the maximum of the amplitude in $\eta_1=0$~cm, on panels (b) and (d). \ADD{As it is defined, the width represents a characteristic scale of the attractor branch, which is different from the wave length. Thus, both can be measured}. The width depends of course on the definition taken, which, however, affects only the constant $C$ in the scaling~(\ref{eq:scaling_width}). 

\subsubsection{Formation and decay of attractors in the linear regime\label{linear_wavelength}}

Experiments~\cite{HBDM2008} and numerical simulations~\cite{GSP2008} have already given the evolution of the wave number of the branches of the attractor as a function of time, using spectra made from a cut through branch 1. These studies have been done during the decay of the attractor, once the forcing has been stopped. Here we demonstrate a typical linear scenario of the transient behavior of the wave attractor. The steady-state scaling is described in more detail in the next paragraph. Experiments have been performed in the large tank in order to follow the \ADD{development of} branch 1 of an attractor during the growth, the stationary state and the decay of the attractor. The amplitude of the wave generator has been changed between the different experiments, to study the effect of forcing on the evolution of the branches. 
Two series of experiments have been performed, using the same filling of the tank. The second series has been carried on the \DEL{next }day \ADD{after}\DEL{of} the first series. \ADD{The stratification profile has been measured at the beginning of the two days and no difference has been found, except those due to the diffusion in the thin mixed layers at the top and the bottom of the fluid. Mixing can be neglected here because the forcing is applied on a short period of time.} This ensures that all experiments have exactly the same geometry, the same stratification and the same forcing frequency. The operational point in $(d,\tau)$-space is fixed at  $(d,\tau)=(0.34,1.81)$.  Experiments of the first series are growth and decay experiments, with different amplitudes of forcing. This means that the wave generator is started at the beginning of the experiment, the attractor grows until it reaches the steady state and then the wave generator is stopped. Data are collected continuously during the three phases, until the fluid in the tank comes back to rest. Experiments of the second series are only growth experiments, with different forcing amplitudes. 
In this section, we discuss only the experimental results obtained for stable attractors. 
Experiments with unstable attractors at large forcing amplitudes are presented in section IV. 

\begin{figure}[h!]
	\begin{center}
		\includegraphics[width=0.7\linewidth,clip=]{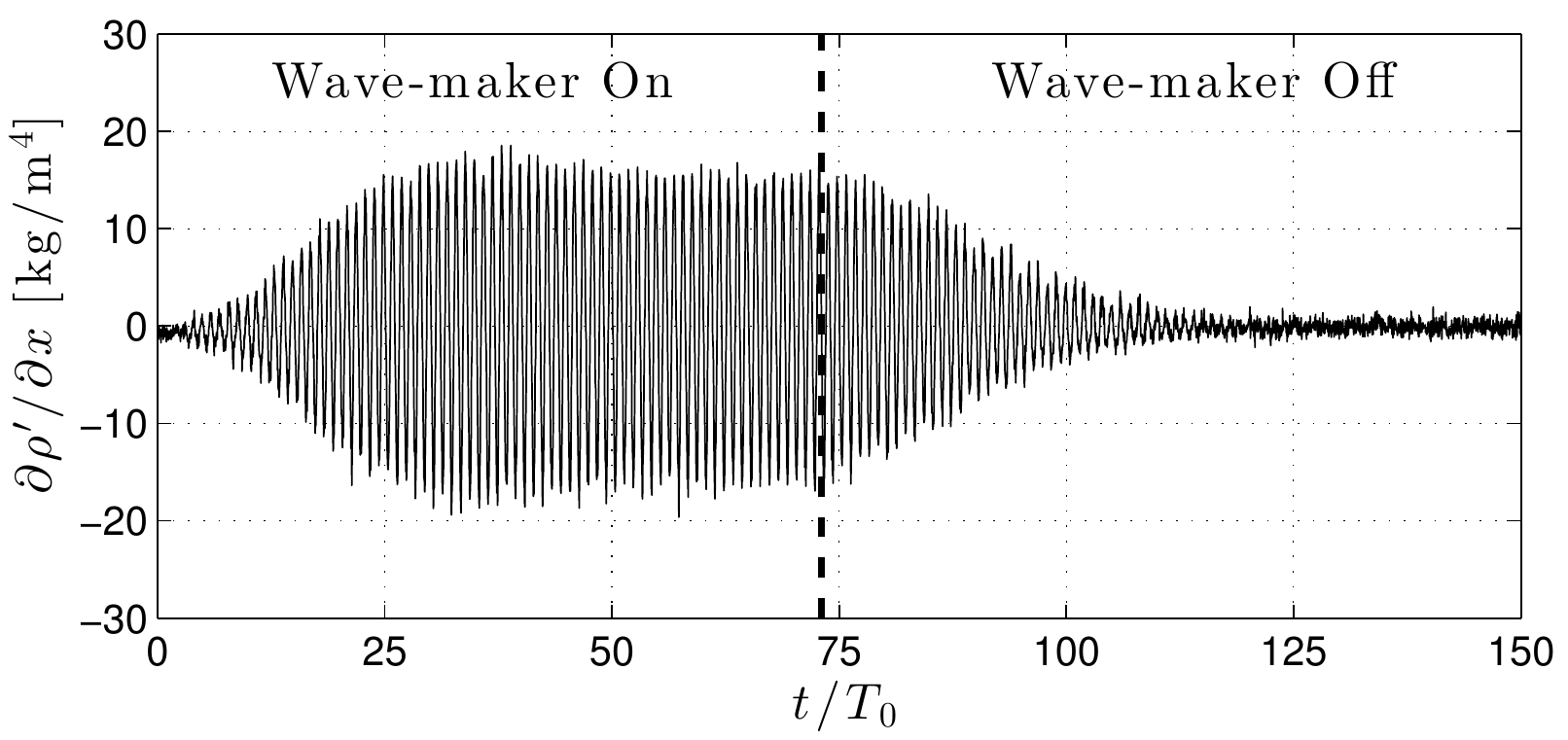}
		\caption{Time history of the horizontal density gradient fields in a point located at $x=85$~cm and $y=63$~cm, on branch 1. The wave generator is started at $t=0~T_0$ and is stopped at $t=73~T_0$.	\label{fig:growth_decay_branch1}}
	\end{center}
\end{figure}

Typical growth, steady state and decay are shown in Fig.~\ref{fig:growth_decay_branch1}, using the time-history of the horizontal density gradient field in a point located on branch 1. \ADD{A similar figure can be found in \cite{HBDM2008}, with less details during the forced stage.} The amplitude starts to grow until the attractor reaches a steady-state regime around $40\,T_0$. This steady-state regime is maintained, until the wave generator is stopped at $t=73\,T_0$. Then, the attractor decays and no motion is observed after $t=115\,T_0$. 

\begin{figure}[h!]
	\begin{center}
		\includegraphics[width=0.65\linewidth,clip=]{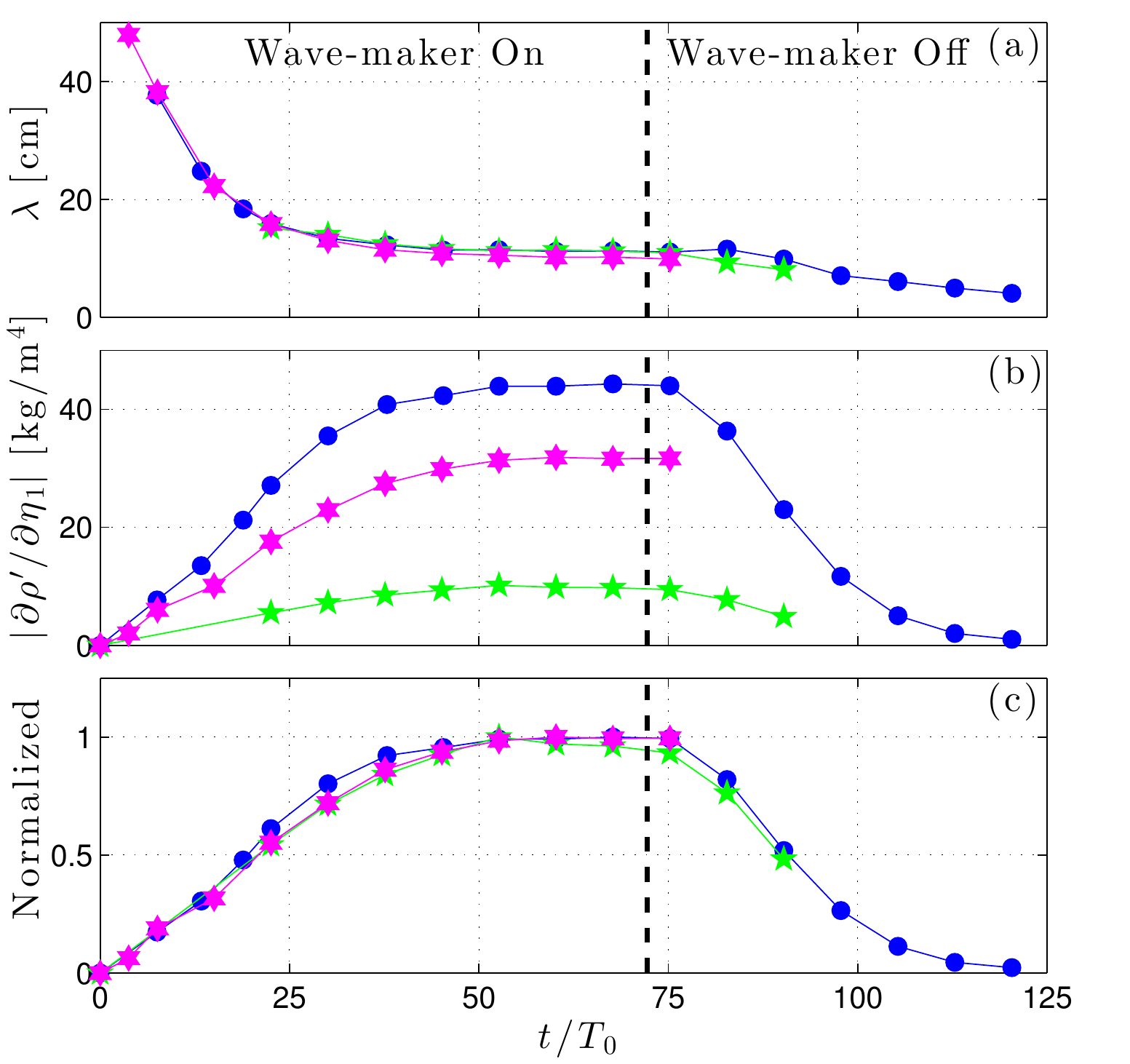}
		\caption{(a): Wavelength of the attractor branch $1$ as a function of time. (b): Transverse density gradient fields amplitude of the attractor branch $1$ as a function of time. (c): \DEL{Normalized t}\ADD{T}ransverse density gradient fields amplitude, \ADD{normalized} with respect to the steady state, as a function of time. The different symbols and colors indicate experiments of different amplitudes: green pentagons for $a=0.7$~mm, magenta hexagons for $a=1.5$~mm and blue dots for $a=2.2$~mm. The symbols show the instants where the wavelength and the amplitude have been measured. There are two growth and decay experiments (green and blue points) and one growth experiment (magenta points). The vertical dashed line represents the moment when the wave generator was stopped.\label{fig:etablissement_attracteur_stable}}
	\end{center}
\end{figure}

In order to measure the peak wave number and width of branch~$1$ as a function of time, the analysis presented in Fig.~\ref{fig:explication_coupe} has been performed on several images, using always the same oblique cut. The complete evolution in time of the wavelength $\lambda$, defined as $\lambda=2\pi/k_{\textrm{peak}}$, is shown in Fig.~\ref{fig:etablissement_attracteur_stable}(a), for three experiments with different amplitudes of the wave generator. The amplitudes are $a=0.7$ (green), $1.5$ (magenta) and $2.1$~mm (blue). The attractors are stable for these amplitudes. One can first note that the attractor wavelength is totally independent of the amplitude of the wave generator at any time, because all the three curves are superimposed. Secondly, one can see how the attractor set-up \DEL{is a machine}\ADD{has the ability} to decrease the wavelengths. Indeed, the wave generator injects a very large vertical wavelength, typically of $184$~cm, which is equal to two times the height of the large tank. While the waves reflect on the slope, the energy is focused and the wavelengths are smaller and smaller until a steady state is reached\ADD{, where}\DEL{. In this state,} the focusing is balanced by the viscous broadening\ADD{~\cite{HBDM2008}}. The ratio between the injected scale and the scale of the attractor in the steady state is found to be around $13.5$. This value is larger than $9$, which is the value of the same ratio reported for a small tank attractor experiment~\cite{SED2013} in the  focusing linear regime. Thus, the large tank allows an energy transfer through a larger range of scales than the small tank. Once the wave generator has been stopped, there is no more large injected wavelength and the slope focuses the waves into smaller and smaller wavelength until all the energy is damped by viscosity\ADD{~\cite{HBDM2008}}.

The influence of the amplitude of the wave generator can be seen in Fig.~\ref{fig:etablissement_attracteur_stable}(b), showing the transverse density gradient field amplitude $|\partial \rho' / \partial \eta_1|$ as a function of time, for the same three different experiments as in~Fig.~\ref{fig:etablissement_attracteur_stable}(a). The larger the amplitude of the wave generator, the larger the amplitude reached by the steady state. Thus, for stable attractors, all the energy emitted by the wave generator is focused into an attractor with the same geometrical characteristics. The only difference is that the amplitude varies with the one of the wave generator. Figure~\ref{fig:etablissement_attracteur_stable}(c) shows the normalized version of the curves presented in Fig.~\ref{fig:etablissement_attracteur_stable}(b). Each curve is now divided by the maximum of amplitude, reached during the steady state. All curves collapse well, showing that the process is purely linear: only the competition between the focusing and viscous broadening matters, while the amplitude of the wave generator affects only the amplitude of the branches. 

\subsubsection{Relation between the beam width and the wavelength\label{width_amplitude_time}}

\begin{figure}[h!]
\begin{center}
  \includegraphics[width=0.7\linewidth,clip=]{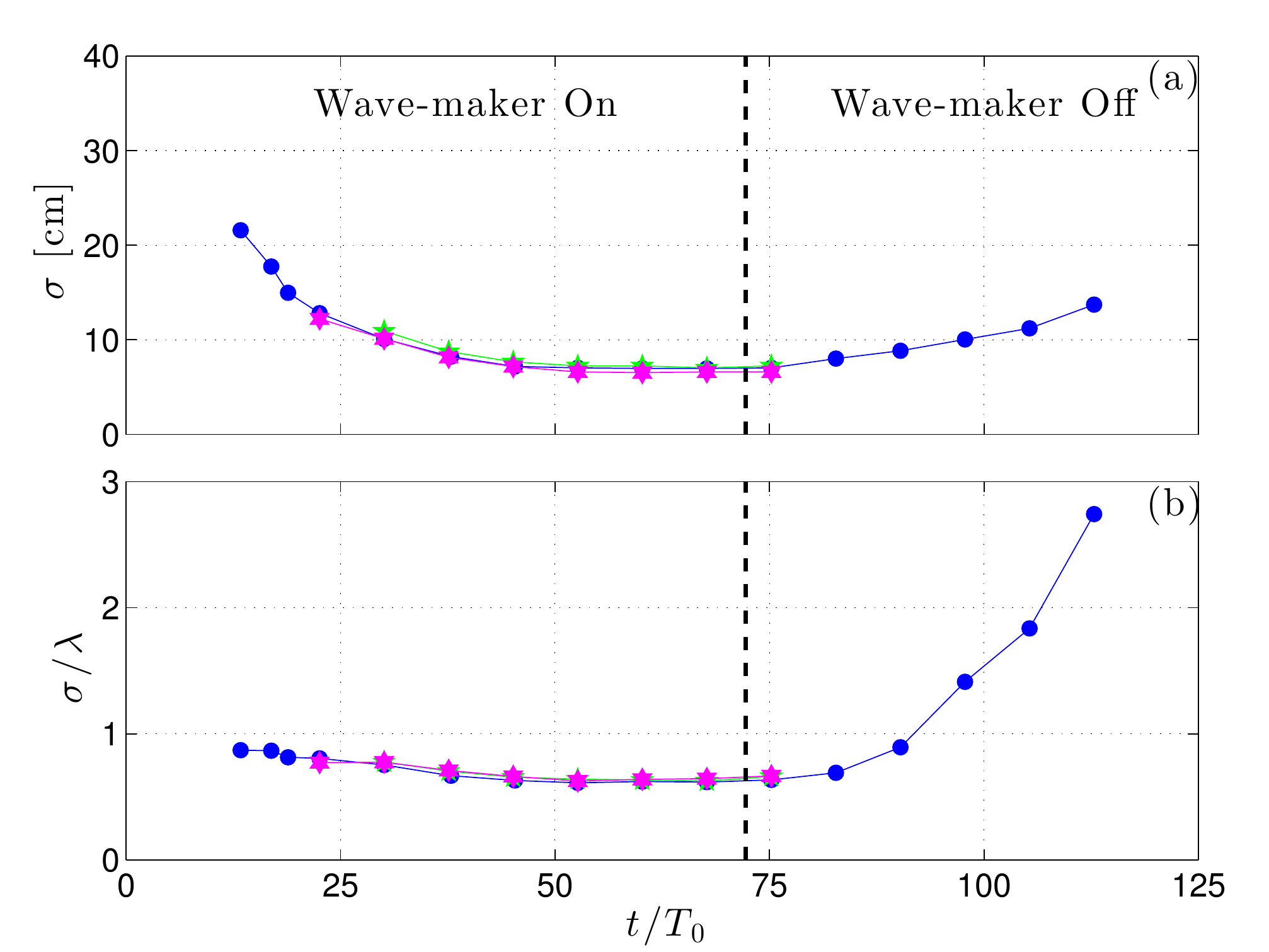}
    \caption{(a): Width $\sigma$ of branch~$1$ as a function of time, for the same three experiments as in figure~\ref{fig:etablissement_attracteur_stable}. (b): Ratio between the width $\sigma$ and the wavelength $\lambda$ of branch~$1$ as a function of time. The dashed black lines show the time where the wave-maker has been stopped. The symbols are the same as the ones in figure~\ref{fig:etablissement_attracteur_stable}.\label{fig:width_evolution}}
   \end{center}
\end{figure}

{Figure~\ref{fig:width_evolution}(a) shows the evolution of the width $\sigma$ of the branch~$1$ as a function of time, for the same three experiments as in Fig.~\ref{fig:etablissement_attracteur_stable}. The curves collapse \ADD{as} well as for the wavelength\DEL{ evolution in time}, presented in Fig.~\ref{fig:etablissement_attracteur_stable}(a). During the growth and the steady state of the attractor, the behavior of the width is very similar to the one of the wavelength. Nevertheless, after the stop of the wave-maker, the width of branch~$1$ increases slightly while the wavelength decreases. The width-over-wavelength ratio gives an idea of the number of wavelengths that are present in the width of the branch. This ratio is plotted, as a function of time, in Fig.~\ref{fig:width_evolution}(b). During the growth, this ratio slightly decays but one can consider that the ratio is more or less constant during the growth and the steady state. This means that there is only one wavelength in the width of the branch during these two phases. This is consistent with Figs.~\ref{fig:explication_coupe}(a) and~(b). Once the wave-maker has been stopped, this ratio increases drastically. Indeed, the wavelength decreases a lot, due to focusing, while the width of the beam slightly increases. This shows that there \ADD{are}\DEL{is} more wavelengths \DEL{that are }present in the width of the branch, as the attractor decays.}

Figure~\ref{fig:width_evolution} highlights that, in the linear regime, the wavelength and the width of the attractor branches have the same behavior, when the attractor is growing or in a steady state. Thus, this indicates that they scale similarly, as given in equation~(\ref{eq:scaling_width}) for the width. \ADD{Indeed, a similar scaling for the wavelength can be found in~\cite{HBDM2008}.}

\subsubsection{Scaling in the linear steady-state regime}

\begin{figure}[t!]
	\begin{center}
		\includegraphics[width=0.9\linewidth,clip=]{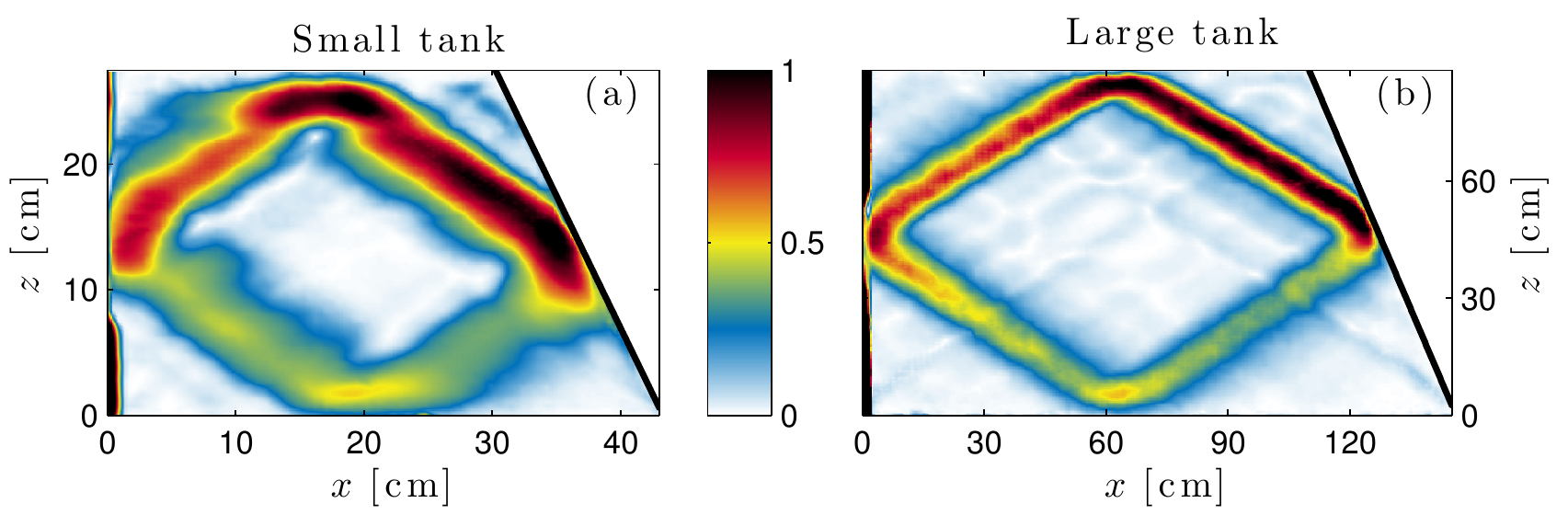}
		\vspace{-0.5cm}
		\caption{$|\partial \rho' / \partial x|$ filtered around $\omega_0$ and normalized by the amplitude of the branch~$1$ for attractors made in the small (a) and large (b) tanks. $(d,\tau)=(0.38,1.85)$ for (a) and $(d,\tau)=(0.52,1.83)$ for~(b).\label{fig:comparison_tank}}
	\end{center}
\end{figure}

Let us now consider in more details the scaling for the width of the attractor in the linear steady-state regime given by equation~(\ref{eq:scaling_width}). Figure~\ref{fig:comparison_tank} shows a comparison between two attractors, one in the small tank \ADD{(left panel)} and the other one in the large tank \ADD{(right panel)}, both observed using SyS. The two attractors have reasonably close $(d,\tau)$ parameters but different scales. The horizontal and vertical scales of the trapezoid in panel (a) are approximately $3$ times smaller than the ones in panel (b). Thus, plotting these two attractors with the same dimension in Fig.~\ref{fig:comparison_tank} is equivalent to normalize them. The horizontal density gradient fields are represented after filtering in frequency around $\omega_0$ and a normalization by the maximal amplitude of the branch~$1$. Therefore the colorbar is the same for the two attractors and lies in the range $[0-1]$. According to equation~(\ref{eq:scaling_width}), the width of the small tank attractor (panel (a)) appears larger than the width of the large tank attractor (panel (b)). The width measurements made by cutting branch~$1$ of both attractors at the same {distance from the virtual point source}, show that the ratio $\sigma/L_p$  is  $4.55 \times 10^{-2}$ for the small tank attractor and $2.25 \times 10^{-2}$ for the large tank attractor. There is nearly a factor~$2$ between these ratios. The model described in~\cite{GSP2008} and given in equation~(\ref{eq:scaling_width}) predicts that $\sigma/L_p\propto L_p^{-2/3}$, with all other parameters being fixed. Therefore, the ratio between $\sigma/L_p$ of the large and small attractors should be equal to the ratio of the perimeters, equal to $3$, to the power $2/3$: $3^{2/3}=2.0800...\approx 2$. Thus, the width measurements are in reasonably good agreement with the model described in~\cite{GSP2008}.
The energy losses in the boundary layers at the longitudinal walls of the test tank introduce a correction to~(\ref{eq:scaling_width}) as discussed in~\cite{BeckebanzeJFM2017}. Leaving the latter issue aside, let us discuss the extrapolation of~(\ref{eq:scaling_width}) to geophysical scales. 
 
If we consider the experimental set-up scaled with the ocean depth, which is around $4000$~m, the perimeter $L_p$ is approximately equal to $10000$~m. In the ocean, the buoyancy frequency is in between $10^{-4}$ and $10^{-3}$~rad/s. Scaling (\ref{eq:scaling_width}) predicts an attractor with the beam width $\sigma$ of a few meters. For a lake of $100$~m depth and a buoyancy frequency of $10^{-3}$~rad/s, the beam width $\sigma$ is less than one meter. 
The confinement of all the energy in such narrow beams does not seem realistic. Thus, one can assume that attractors in oceans or lakes should obey a different scaling, where nonlinearities play an important role. This is discussed in the next section. \ADD{Note that however, in the ocean, the depth is often much smaller than the length. Thus, the aspect ratio of the attractor may change from the one considered in the paper. This may not be an issue since the parameter $\tau$ remains unchanged and can accommodate a large change of the aspect ratio by allowing for a matching decrease in $\Omega_0=\omega_0/N$. For  astrophysical systems, an aspect ratio close to $1$ as in the attractors of this paper is more relevant. On the contrary, the friction at lateral walls does matter in the experiment: the beams are therefore slightly wider than they would be in a pure 2D case without lateral walls. It means that if we extrapolate the experimental results in a “thought experiment” to a larger scale, where the lateral walls are not relevant, we get a somewhat overestimated width.}

\section{Scaling in unstable attractors: a signature of triadic resonance instability}
\label{Scaling in unstable attractors: a signature of triadic resonance instability}

\subsection{Theoretical preliminaries}

The mechanism of instability in wave attractors is similar to the classic\ADD{al} concept of triadic resonance instability (TRI)~\cite{SED2013}. TRI is best studied for the idealized case, with a monochromatic in time and space carrier wave as a basic state which feeds two secondary waves via nonlinear resonant interactions. The resonance occurs when temporal condition for frequencies 
\begin{equation}\Omega_{1}+\Omega_{2}=\Omega_{0}\end{equation}
and spatial condition for wave vectors 
\begin{equation}\vec{k}_{1}+\vec{k}_{2}=\vec{k}_{0}\end{equation}
are satisfied, where subscripts 0, 1 and 2 refer to the primary, and two secondary waves, respectively. In a wave attractor, the wave beams serve as a primary wave, and the resonance conditions are satisfied with a good accuracy~\cite{SED2013}, thereby providing a consistent physical framework to the observed phenomena. The \DEL{onset of the }resonance is \DEL{similar to }\ADD{thus governed by the classical concepts of TRI}~\cite{KoudellaStaquet2006,BDJO2013}, \ADD{but} with the effect of finite wave-beam width involved~\citep{KarimiAkylas2014,BSDBOJ2014}.
The latter is important since the subharmonic waves can serve as an energy sink only if they can extract substantial energy from the primary wave before leaving the beam~\cite{BSDBOJ2014,Dauxoisetal2018}. If the energy injection into the wave attractor is large, secondary waves can reach a large amplitude and therefore be also unstable on a faster time scale, generating a cascade of triadic interactions transferring energy to small scales where the viscous dissipation becomes significant. Since secondary waves provide a more powerful mechanism of momentum flux from the primary wave than the molecular viscosity alone, one can expect,  in the nonlinear regime \ADD{and particularly in natural systems}, the broadening of wave beams in attractors. In that aspect, the role of the cascade of triadic interactions is conceptually similar to the role of turbulent viscosity. The broadening of the wave beams for a larger energy input is observed in numerical calculations for inertial wave attractors performed in~\cite{JouveOgilvie2014}. In  experiments performed with a small set-up~\cite{SED2013,BrouzetEPL2016,BrouzetJFM2016,BrouzetJFM2017}, the effect is weak due to viscous effects~\cite{BeckebanzeJFM2017}. However, it is fully present in the upscaled set-up used in the present work. Below, we describe quantitative measurements of the length of the primary wave in internal wave attractors in the nonlinear regime. Additionally, we perform 2D numerical simulations demonstrating the universality of results in the studied range of parameters.

\subsection{Experiments with unstable attractors\label{branch_size_TRI}}%%

For stable attractors, the size of the beam is directly linked with the geometry and the focusing, as shown in the previous section. Below, we reveal the effect of forcing on the scaling of the beam wavelength and wave amplitude. As in the previous section, the experiments are carried out in the large tank, with the SyS visualization technique, with systematic variation of the wave generator amplitude $a$ beyond the stability threshold. The geometrical parameters and stratification are fixed so that $(d,\tau)=(0.34,1.81)$. {Note that the curves of the beam width of unstable attractors are not shown here but, as for stable attractors, they are similar to the ones of the wavelength.}

\begin{figure}[h!]
\begin{center}
    \includegraphics[width=0.55\linewidth,clip=]{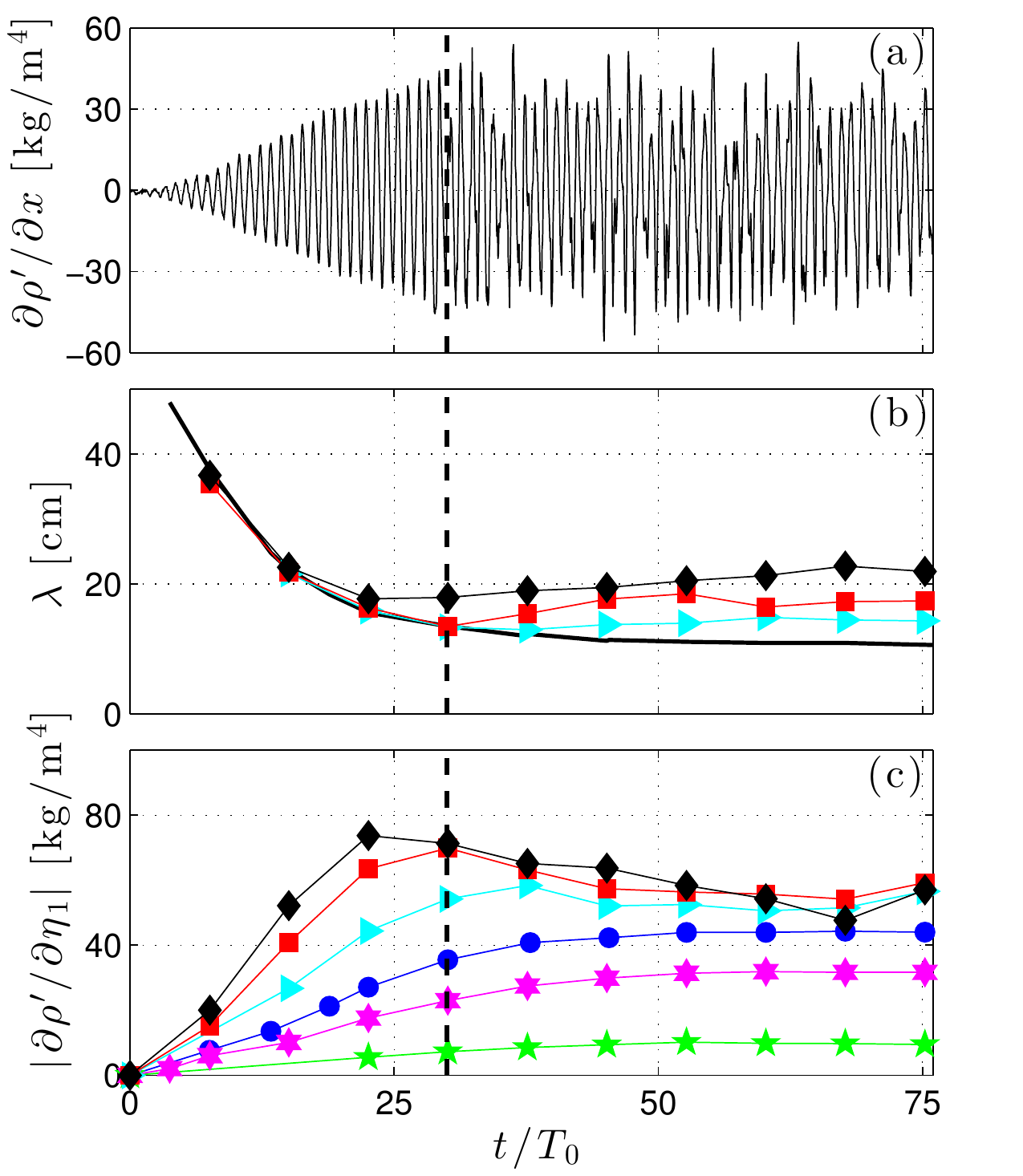}
   \caption{(a): Time-history of the horizontal density gradient field at one point located on branch~$1$ for the unstable attractor with $a=4.4$~mm (Exp.~$6$ of Table~\ref{tab:series_exp}). (b) and (c): Wavelength and amplitude of branch $1$ as a function of time for three stable and three unstable attractors (see Table~\ref{tab:series_exp}). The three curves for the stable attractors are the ones presented in Fig.~\ref{fig:etablissement_attracteur_stable}. On panel (b), for the sake of clarity, the average of the wavelengths of these three stable experiments is plotted as a unique solid black line.
The vertical dashed line on the three panels represents the time corresponding to visible onset of TRI in panel~(a), which corresponds to red squares in panels~(b) and (c). }   \label{fig:wave_length_TRI}
\end{center}
\end{figure}

The wavelength $\lambda$ and the amplitude $|\partial \rho' / \partial \eta_1|$ of branch~$1$  are measured as described in section~\ref{branch_wavelength_width}, after frequency and space filtering. Thus, the presence of the TRI does not disturb the measurements of the wavelength and the amplitude of the primary wave (attractor beam) because the wave fields oscillating at other frequencies than $\Omega_0$ do not appear in the filtered data. Figure~\ref{fig:wave_length_TRI} shows the time history of the horizontal density gradient field (a) at a point  located on the branch $1$  for the unstable attractor with $a=4.4$~mm (see Exp.~$6$ in Table~\ref{tab:series_exp}) and the time-histories of wavelength (b) and wave amplitude (c) at different values of $a$. 
Since, in the linear regime, the data for the wavelengths collapse onto a single curve (see Fig.~\ref{fig:etablissement_attracteur_stable}), for the sake of clarity, we plot the average of these data as a function of time in Fig.~\ref{fig:wave_length_TRI}(b), using a solid black line. 
Let us focus on Exp.~6 of Table~\ref{tab:series_exp}, plotted using red squares. One can see that the wavelength of this unstable attractor follows the universal curve until $30\,T_0$, the time corresponding to the onset of TRI in panel (a). This means that, before the start of the instability, the attractor experiences the linear geometric focusing following the scenario described in previous section. After $30\,T_0$, the wavelength of the unstable attractor departs from the universal curve for linear regime (solid black line). In Fig.~\ref{fig:wave_length_TRI}(c), where the amplitude of branch $1$ is plotted as a function of time, one can see that the amplitude of the attractor reaches a maximum around $30\,T_0$, when the instability starts. Thus, through linear focusing, all the energy injected by the wave generator is focused into the attractor and the amplitude increases until the TRI threshold is reached. After $t=30\,T_0$, when the amount of energy focused into the branch~$1$ is too high, the instability starts. This brings the attractor to a larger wavelength, which appears constant with time beyond a transient growth, after roughly $50\,T_0$. The amplitude of the attractor decays until it reaches a plateau, around $50\,T_0$. The duration of transients for the wavelength and amplitude after the onset of TRI is nearly the same.  Thus, TRI balances the geometrical focusing by transferring a part of energy of the primary wave into the secondary waves. Consequently, the width of the branch $1$ increases as compared to the linear case. 

The two other unstable experiments (Exps.~$4$ and $8$ with cyan triangles and black diamonds) exhibit a behavior similar to Exp.~$6$ (red squares). As can be seen in Fig.~\ref{fig:wave_length_TRI}(b), the larger the amplitude of the wave generator, the earlier the wavelength curve departs from the linear scenario and the earlier the maximum of the amplitude is reached. For the fully developed nonlinear regime, i.e. after the transient following the onset of TRI, the larger the amplitude of the wave generator, the larger the wavelength of the primary wave. In the studied range of parameters, the amplitudes of the unstable attractors seem to reach saturation around the same value, independent of the amplitude of the wave generator as seen in~\ref{fig:wave_length_TRI}(c). After the transient, the final values of the wavelength and the amplitude are respectively denoted $\lambda_f$ and $|\partial \rho' / \partial \eta_1|_f$. 

\begin{figure}[h!]
\begin{center}
\vspace{0.45cm}
  \includegraphics[width=0.85\linewidth,clip=]{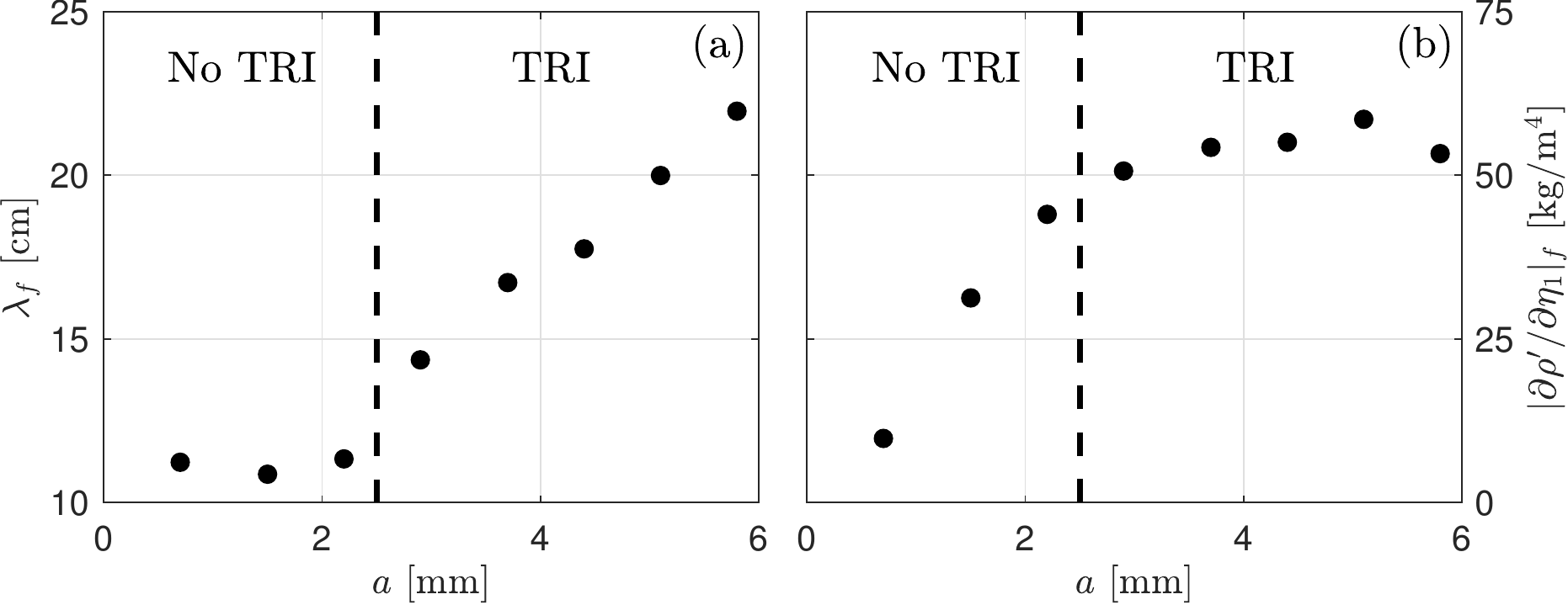}
  \vspace{-0.25cm}
   \caption{Final wavelength (a) and amplitude (b) 
    of the branch~$1$ of attractors as a function of the amplitude of the wave generator~$a$. The dashed lines on the two 
     panels delimit the lowest amplitudes, where no TRI is observed, and the highest amplitudes, where TRI is observed. 
 }   \label{fig:psi_VS_a}
\end{center}
\end{figure}

The final values for the wavelengths and the amplitudes are plotted as a function of the amplitude of the wave generator $a$ in Figs.~\ref{fig:psi_VS_a}(a) and (b). The values are determined by averaging the wavelengths or the amplitudes between $50$ and $75\,T_0$. For Fig.~\ref{fig:psi_VS_a}, the eight experiments of Table~\ref{tab:series_exp} (three stable and five unstable) have been used. Among the five unstable ones, only three have been plotted in Fig.~\ref{fig:wave_length_TRI}, for the sake of clarity (see Table~\ref{tab:series_exp}). Nevertheless, the extra-two unstable experiments (numbered $5$ and $7$) exhibit very similar characteristics as the ones presented in Fig.~\ref{fig:wave_length_TRI}. Figures~\ref{fig:psi_VS_a}(a) and (b) summarize the steady states reached by the attractors. When the amplitude of the wave generator is low, there is no TRI: the wavelength is constant and independent of $a$ while the amplitude increases with $a$. When the amplitude of the wave generator is large, TRI appears: the wavelength increases with $a$ while the amplitude is constant and independent of $a$. Note that for the upper bound of the range of amplitude studied in experiments the value of $\lambda_{f}$ is roughly twice higher than at the lower bound. Laminar scaling \eqref{eq:scaling_width} suggest that the width of the wave beam is proportional to~$\nu^{1/3}$. Thus, a comparable increase of the beam width in linear case would require an artificial ``turbulent" viscosity $8$ times higher than the molecular one. 

\subsection{Numerical simulations in 2D setting\label{numerics}}

\begin{figure}[b!]
\begin{center}
	\includegraphics[width=0.85\linewidth,clip=]{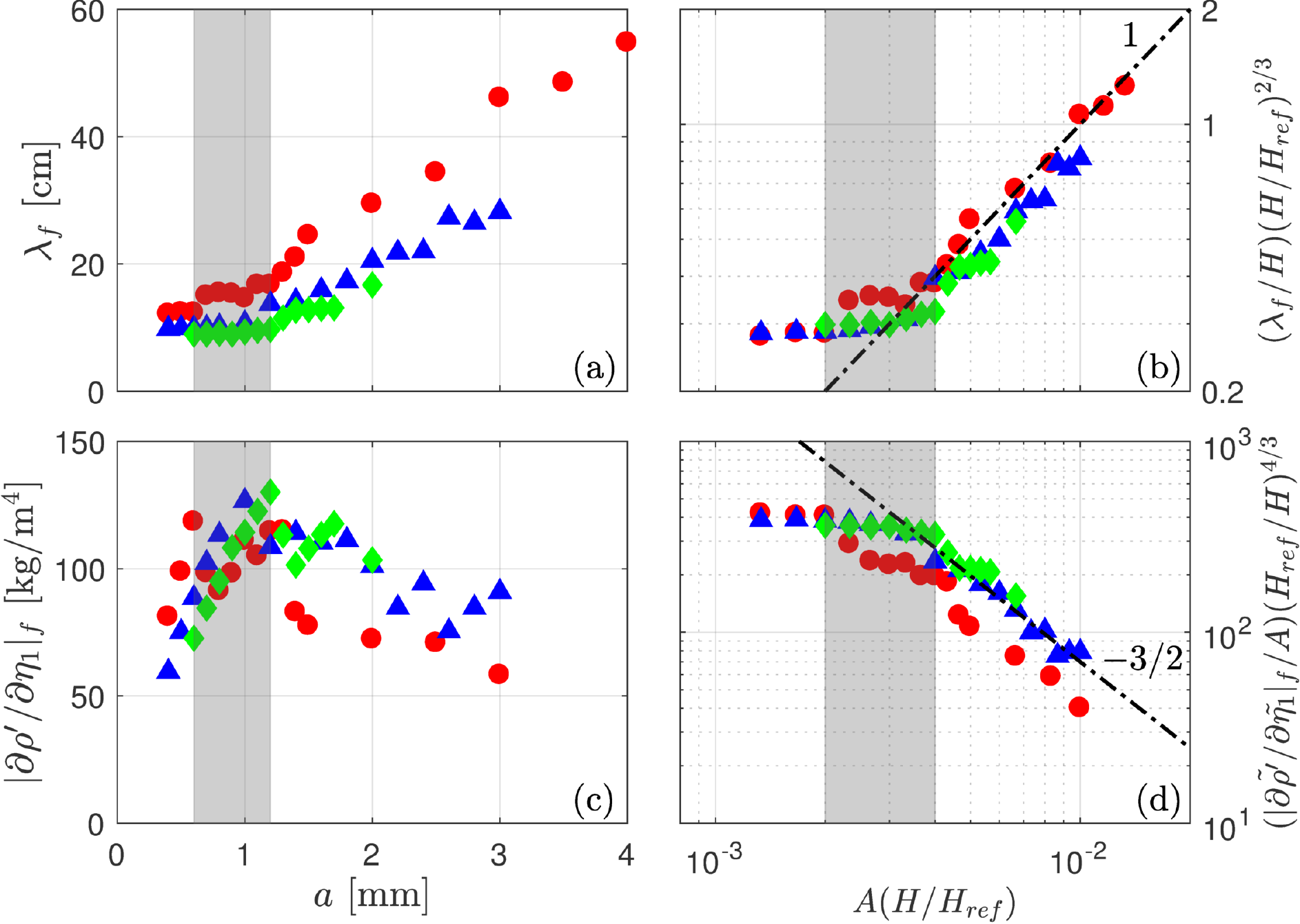}
	\caption{\DEL{Non-dimensional wavelength $\lambda_{f}/H$ (a) and ratio of the non-dimensional density gradient $|\partial \tilde{\rho'} / \partial \tilde{\eta_1}|_f$ and the non-dimensional amplitude of wave generator $A=a/H$ (c) as a function of non-dimensional amplitude of wave generator $A=a/H$}\ADD{Final wavelength (a) and amplitude (c) as a function of the amplitude of the wave generator} for three different values of the water depth: $H=92$~cm (red circles), $H=46$~cm (blue triangles) and $H=30$~cm (green diamonds). Panel (b) shows on log-log scale the data from panel (a), with the ordinate rescaled as $(\lambda_{f}/H)(H/H_{ref})^{2/3}$ and the abscisse rescaled as $A\times (H/H_{ref})$, where $H_{ref}=30$~cm is taken as reference value. A black dashed \ADD{dotted} line with slope $1$ is drawn to guide the eye. Panel (d) shows on log-log scale the data from panel (c), with the ordinate rescaled as $(|\partial \tilde{\rho'} / \partial \tilde{\eta_1}|_f/A)(H_{ref}/H)^{4/3}$ and the abscisse rescaled as $A\times (H/H_{ref})$. Again, $H_{ref}=30$~cm is taken as reference value. The black dashed dotted line indicates a $-3/2$ slope. \ADD{Light gray stripes in the four panels show approximatively the transition between stable and unstable attractors, where the left (respectively right) borders of the stripes are relevant to the larger (resp. smaller) set-ups.}}\label{fig:2Dnumerics}
\end{center}
\end{figure}

As already mentioned, the scaling for the beam width in experiments is obscured by the effect of viscous losses at lateral walls~\cite{BeckebanzeJFM2017}. The numerical calculations presented in~\cite{BrouzetJFM2016} show that in a test tank of width $17$~cm and working depth $30$~cm about 25\% of energy dissipation occur in boundary layers at lateral walls. To clarify the scaling laws in linear and nonlinear regimes, we performed a series of numerical calculations in a purely 2D setting for geometrically similar set-ups \DEL{which}\ADD{that} are characterized by $H/L=0.625$, $\alpha=27^{\circ}$, $\Omega=0.575$, $N=0.822$~rad/s. \ADD{To extend the \DEL{too few }experimental results discussed in previous subsection, w}e consider three values of the water depth $H=30$, $46$ and $92$~cm, and a range of amplitudes of the wave generator $a$ from $0.4$ to $4$~mm. \DEL{Here we introduce the non-dimensional amplitude of wave generator $A=a/H$. }Figure~\ref{fig:2Dnumerics} presents the numerical data for the \ADD{final wavelength $\lambda_{f}$ in panel (a) and the final amplitude of the density gradient $|\partial \rho' / \partial \eta_1|_f$ in panel (c) as a function of the amplitude of the wave generator~$a$. This allows us a direct comparison with the experimental data, presented similarly in Fig.~\ref{fig:psi_VS_a}. It can be seen that the wavelength remains constant at low forcing, and starts to grow when a given critical value of forcing is reached. This critical forcing amplitude seems to depend on the different water depths and belongs to the light gray stripes represented in the four panels of Fig.~\ref{fig:2Dnumerics}. The density gradient grows almost linearly with the forcing amplitude in the linear regime until it reaches a critical value, around $130$~kg$/$m$^4$ in the three different simulations. In the non-linear regime, the density gradient first saturates before decreasing for the largest forcing amplitudes.}

\ADD{One can remark that TRI starts for a smaller forcing amplitude in the simulations (in between $a=0.6$~and $1.2$~mm) than in the experiments (around $a=2.5$~mm). This is due to the damping of the lateral walls in the experimental setup. Thus, experiments explore a smaller forcing range than numerical simulations. Figures~\ref{fig:psi_VS_a}(b) and~\ref{fig:2Dnumerics}(c) show a critical density gradient around $55$~kg$/$m$^4$ in the experiment and around $130$~kg$/$m$^4$ in the three different simulations. This can be associated with a critical "wave steepness", that triggers the TRI. Experimentally, only the saturation at the critical "steepness" is observed while for simulations, a decay is present for very large forcing. This decay is not captured by the experiments because the forcing range is limited. Thus, the experimental and numerical trends are in good agreement}.

\ADD{A rescaled version of data from \DEL{panel}\ADD{Fig.~\ref{fig:2Dnumerics}}(a) and~(c) is shown in \DEL{panel}\ADD{Fig.~\ref{fig:2Dnumerics}}(b) and~(d) on log-log scale. Here we introduce the non-dimensional amplitude of wave generator $A=a/H$, the} non-dimensional wavelength $\lambda_{f}/H$ in \DEL{panel (a)}\ADD{Fig.~\ref{fig:2Dnumerics}(b)} and the ratio between the non-dimensional density gradient 
\begin{equation}
|\partial \tilde{\rho'} / \partial \tilde{\eta_1}|_f\equiv \frac{|\partial \rho' / \partial \eta_1|_f }{\bar{\rho}N^2/g}
\end{equation} 
and the dimensionless amplitude of wave generator~$A$ in \DEL{panel (c)}\ADD{Fig.~\ref{fig:2Dnumerics}(d).}\DEL{ as a function of the dimensionless amplitude of the wave generator~$A$. It can be seen that the wavelength remains constant at low forcing, and starts to grow when a given critical value of forcing is reached. The ratio between the non-dimensional density gradient and the dimensionless amplitude of wave generator $A$ remains constant at low forcing, \DEL{what}\ADD{which} corresponds to a linear growth for the dimensional density gradient $|\partial \rho' / \partial \eta_1|_f $, and \DEL{falls down}\ADD{decreases} in the nonlinear regime. These trends are in  full agreement with the experimental behavior shown in Fig.~\ref{fig:psi_VS_a}.} \DEL{A rescaled version of data from \DEL{panel}\ADD{Fig.~\ref{fig:2Dnumerics}}(a) is shown in \DEL{panel}\ADD{Fig.~\ref{fig:2Dnumerics}}(b) on log-log scale.}\ADD{These three quantities are combined using the ratio $H/H_{ref}$, where $H_{ref}=30$~cm.} It can be seen that all the data collapse reasonably well on a common master curve, which has a horizontal branch corresponding to the linear scaling \eqref{eq:scaling_width} and a sloping branch corresponding to the nonlinear scaling due to the onset of TRI. \DEL{\DEL{Panel}\ADD{Figure~\ref{fig:2Dnumerics}}(d) shows also a rescaled version of the data in \DEL{panel}\ADD{Fig.~\ref{fig:2Dnumerics}}(c) on log-log scale, using a similar rescaling as in \DEL{panel}\ADD{Fig.~\ref{fig:2Dnumerics}}(b). Again, all data collapse correctly on a master curve with two different regimes: the linear one with a plateau at low forcing and the nonlinear one with a negative slope for higher forcing.} Therefore, in the studied range of fluid depth which corresponds to typical values reached in experimental facilities, the behavior of wave attractors is universal, and the critical value of forcing corresponding to the transition from linear to nonlinear regime for geometrically similar configurations can be found by rescaling the data from a single experiment. It is worth noting that, in the nonlinear regime, the width of the wave beams increases roughly linearly with the amplitude of forcing, \DEL{what}\ADD{which} implies that an extrapolation of the linear scaling to the nonlinear regime would require an artificial turbulent viscosity proportional to the cube of the forcing amplitude. This illustrates the efficiency of TRI in the transfer of momentum from the primary wave beam.  

\DEL{There are reasons to believe}\ADD{One can expect} that in sufficiently wide experimental tanks, where the effect of viscous losses at lateral walls in the total energy balance is negligible, the results of our 2D calculations should be in a good quantitative agreement with the experimental data. There is a possibility that, in large-scale experiments or at natural conditions, one may observe the development of a 3D instability similar to~\cite{GayenSarkar2010}, with \DEL{certain}\ADD{possible} implications to scaling. However, this issue is beyond the scope of the present study.    

\section{Conclusions}
\label{Conclusion}

In absence of dissipation \ADD{and instability}, internal (inertial) wave attractors in ideal fluids stratified in density (angular momentum) exhibit singular behavior. In particular, internal wave attractors in 2D problem represent closed linear loops, where the energy density is infinite. The regularization of this problem has been based on the concept of interior shear layers, which removes the singularity and yields the scaling for the equilibrium width of the attractor beams~\cite{RGV2001,RVG2002,HBDM2008,GSP2008}. In the linear viscous model, this scaling predicts that the beam width is proportional to the $1/3$ power of the kinematic viscosity and the distance along the beam. Importantly, the equilibrium beam width corresponds to the balance between injection and dissipation of energy in the confined fluid domain. Therefore, the beam width is sensitive to the particular dissipative mechanism operating in the system. 

In the present paper, we report experimental and numerical results on the linear and nonlinear scaling in wave attractors. We use two geometrically similar experimental set-ups, scaled to factor 3, with a classic\ADD{al} trapezoidal geometry of the fluid domain filled with a uniformly stratified fluid. In the linear regime, we recover the classic\ADD{al} viscous scaling for the beam width, which is not affected by variations of the amplitude of the input perturbation. Note that the viscous scaling assumed by the concept of interior shear layers is not exact in a typical experimental set-up due to energy losses at lateral walls as discussed in~\cite{BeckebanzeJFM2017}. As the input perturbation increases beyond a given threshold, we observe the onset of triadic resonance instability, which replaces the viscous transfer of momentum by a more efficient mechanism involving the flux of momentum due to secondary waves emanating from the primary wave beam. In this nonlinear regime, the beam width increases linearly with the amplitude of the input perturbation. The growth of the beam width can be qualitatively interpreted as the effect of fictitious ``turbulent" viscosity which increases as the forcing amplitude with a power $3$. 

Numerical 2D simulations performed in the present study yield a similar behavior, which is also in qualitative agreement with an apparent broadening of wave beams at large forcing in inertial wave attractors~\cite{JouveOgilvie2014}. The numerical simulations have been performed for three geometrically similar set-ups, in the range of fluid depth between $0.3$ and $1$~m corresponding to typical experimental conditions. We show that under the appropriate scaling, the results for the beam width are universal with a reasonable accuracy both in linear and nonlinear regimes. 

Future research can be pursued in several directions. At large scales, there is a yet unexplored possibility of 3D instability in a nominally 2D problem~\cite{GayenSarkar2010}. At strong forcing, some mixing in wave attractors may occur as described in~\cite{BrouzetEPL2016,BrouzetJFM2017}, motivating further research on scaling of ``turbulent" vertical diffusivity for highly nonlinear regimes. Exploration of highly nonlinear regimes  for wave attractors in spherical shells~\cite{MaasHarlander2007,RieutordValdettaro2010,BaruteauRieutord2013} may also present an interesting line for further research.

\begin{acknowledgments}
This work was supported by the LABEX iMUST (ANR-10-LABX-0064) of Universit\'e de Lyon, within 
the program "Investissements d'Avenir" (ANR-11-IDEX-0007) operated by the French National 
Research Agency (ANR), and also supported by Ministry of Education and Science of Russia
(RFMEFI60714X0090, grant number 14.607.21.0090). DNS were
performed on the supercomputer Lomonosov of Moscow State University.
This work has been achieved thanks to the resources of PSMN from ENS de Lyon. E. V. E. gratefully acknowledges \DEL{his appointment}\ADD{support} as a Marie Curie incoming fellow at Laboratoire de physique \ADD{at} ENS de Lyon\DEL{, and a subsequent partial support from IMC NSU}.

\end{acknowledgments}

\end{document}